\documentclass[pra,aps,twocolumn,amsmath,amssymb]{revtex4}
\usepackage{graphicx}
\pdfoutput=1
\usepackage{amsmath}
\usepackage{color}
\usepackage{float}
\setcitestyle{super}
\usepackage[english]{babel}
\usepackage{color,soul}
\usepackage{xcolor}

\begin{document}
\title{Multi-band superconductivity driven by a site-selective mechanism in Mo$_8$Ga$_{41}$ }

\author{Anshu Sirohi$^1$, Surabhi Saha$^2$, Prakriti Neha$^3$, Shekhar Das$^1$, Satyabrata Patnaik$^3$, Tanmoy Das$^2$}

\author{Goutam Sheet$^1$}
\email{goutam@iisermohali.ac.in}

\affiliation{Department of Physical Sciences, Indian Institute of Science Education and Research(IISER) Mohali, Sector 81, S. A. S. Nagar, Manauli, PO: 140306, India.}
\affiliation{$^2$Department of Physics, Indian Institute of Science, Bangalore 560012, India}

\affiliation{$^3$School of Physical Sciences, Jawaharlal Nehru University, New Delhi, PO: 110067, India}

\begin{abstract}

\textbf{The family of the endohedral gallide cluster compounds recently emerged as a new family of superconductors which is expected to host systems displaying unconventional physics. Mo$_8$Ga$_{41}$ is an important member of this family which shows relatively large $T_c \sim$ 10 K and has shown indications of strong electron-phonon coupling and multi-band superconductivity. Here, through direct measurement of superconducting energy gap by scanning tunneling spectroscopy (STS) we demonstrate the existence of two distinct superconducting gaps of magnitude 0.85 meV and 1.6 meV respectively in Mo$_8$Ga$_{41}$. Both the gaps are seen to be conventional in nature as they evolve systematically with temperature as per the predictions of BCS theory. Our band structure calculations reveal that only two specific Mo sites in an unit cell contribute to superconductivity where only $d_{xz}$/$d_{yz}$ and $d_{x^2-y^2}$ orbitals have strong contributions. Our analysis indicates that the site-elective contribution governs the two-gap nature of superconductivity in Mo$_8$Ga$_{41}$.} 

\end{abstract}

\maketitle

The rule of Matthias for predicting new superconductors with higher critical temperatures\cite{Matthias} ($T_c$) that says a higher density of states at the Fermi energy ($E_F$) is expected to lead to a higher $T_c$ is partly followed by the endohedral gallide cluster family of superconductors with lower valence electron counts. For higher electron counts, beyond Mo$_8$Ga$_{41}$ that superconducts below $T_c$ $\sim$ 10 K,\cite{Muller} the architecture of the cluster packing starts playing a dominant role in deciding $T_c$ and in this regime, the $T_c$ goes down though DOS at $E_F$ goes up. This competition makes the $T_c$ of Mo$_8$Ga$_{41}$ maximum in the family.\cite{Cava} Based on clear understanding of the relationship of the superconducting properties with the structural and electronic properties of these coumpounds, new electron counting rules were developed using which it has been possible to predict new superconductors belonging to the family that have also been experimentally realized.\cite{Cava} Like many other family of superconductors, the superconductivity in gallium cluster family might also emerge through complex pairing mechanism and host unconventional physics. Indications of unconventional superconductivity has already been obtained on a number of compounds belonging to this family. For example, in case of superconducting PuCoGa$_5$ it was argued that, like in high T$_c$ cuprates, antiferromagnetic fluctuations might lead to superconducting pairing.\cite{JOP_Bauer, Nature_Sarao, Daghero}

Based on a number of experiments that were employed to study the superconducting phase of Mo$_8$Ga$_{41}$, it was shown that this compound manifests unusually high electron-phonon coupling leading to a large $\Delta/k_BT_c$ ratio and indication of multi-gap superconductivity was also found.\cite{Shevelko_MoGa1, Shevelko_MoGa2} In this Letter, from direct measurement of the superconducting energy gap through scanning tunneling spectroscopy (STS) we show that Mo$_8$Ga$_{41}$, like MgB$_2$, is a two-gap superconductor.\cite{Silva_MgB2, MgB2_2, MgB2_3, MgB2_4, MgB2_5, MgB2_6, MgB2_7, MgB2_8} Two distinct gap structures are clearly resolved in the quasiparticle energy spectra. From detailed temperature dependent experiments we conclude that both the gaps are conventional in nature. Through the band structure calculations, we have also identified the bands responsible for the two respective gaps. Our magnetic field dependent STS experiments further suggest that the interband coupling is weak in Mo$_8$Ga$_{41}$, as in case of MgB$_2$.\cite{Silva_MgB2}    

Compact samples of Mo$_8$Ga$_{41}$ were synthesized through solid state reaction by mixing constituent elements Mo (99.999\%) powder and Ga (99.999\%) pieces in stoichiometric ratio in a quartz ampoule which was evacuated down to 10$^{-4}$ mbar and heated to 850$^o$C and then cooled down very slowly. The samples appeared shiny grey and the formation of Mo$_8$Ga$_{41}$ in single phase was confirmed by powder X-ray diffraction followed by Rietveld analysis. A superconducting transition at $T_c \sim$ 10 K was found from both temperature dependent resistivity and magnetization experiments. The STM and STS experiments were carried out in an ultra-high vacuum (UHV) cryostat working down to 300 mK (Unisoku system with RHK R9 controller). The STM is equipped with a UHV sample preparation chamber, where few layers of the surface was first removed by mild sputtering in an argon environment prior to the STS experiments. This ensured that we probed the pristine surface of Mo$_8$Ga$_{41}$. 
\begin{figure}[h!]
	\centering
		\includegraphics[width=.5\textwidth]{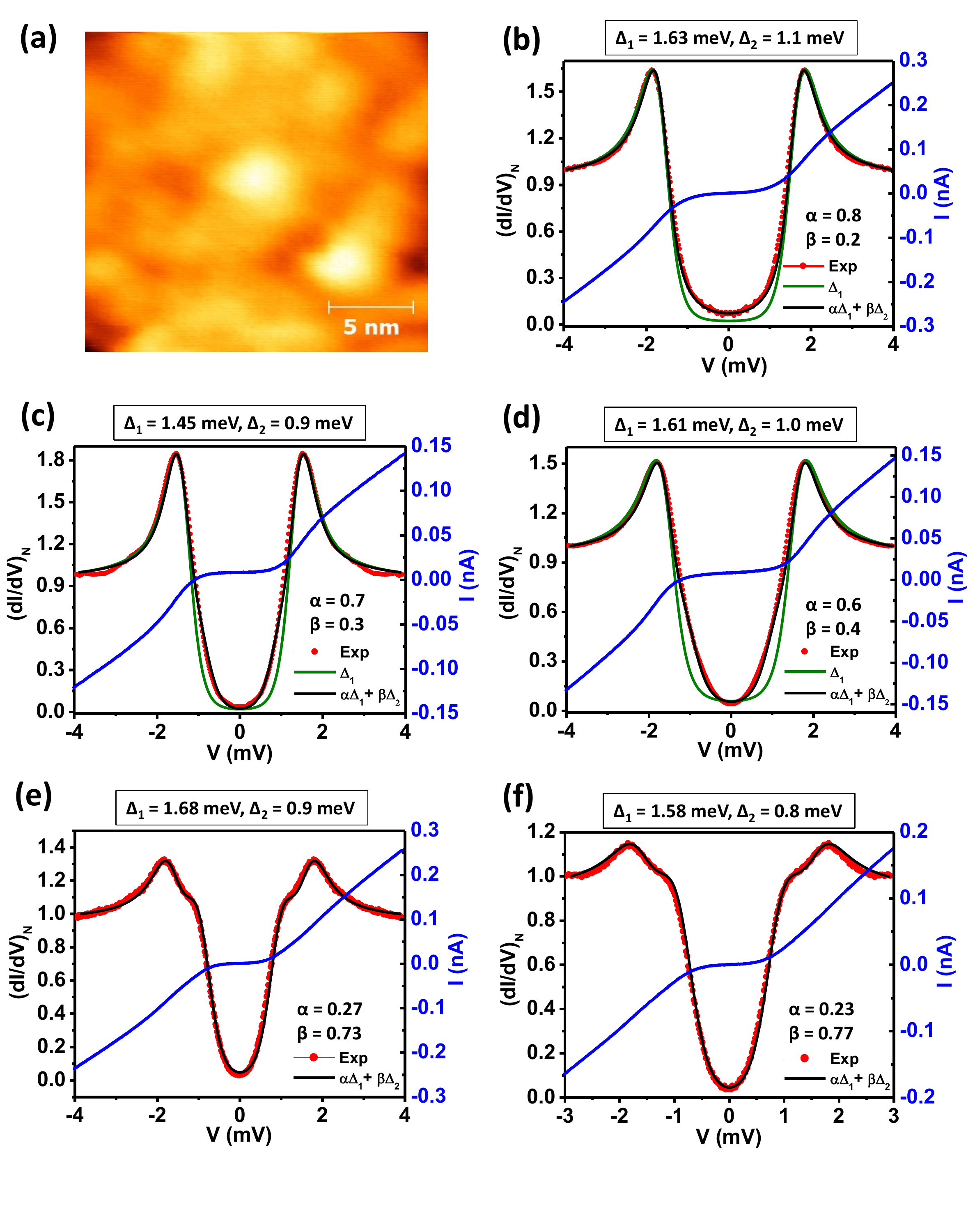}
	\caption{(a) STM topograph image of the sample. (b) – (e) Tunneling spectra ($dI/dV$ vs $V$ plots) with theoretical fits using Dynes equation showing two gaps measured at 1.9 K. The color dots are experimental data, solid lines show fits with single gap (Green line) and with double gap (Black line).}
	\label{Figure 4}
\end{figure}

\begin{figure}[h!]
	\centering
		\includegraphics[width=.44\textwidth]{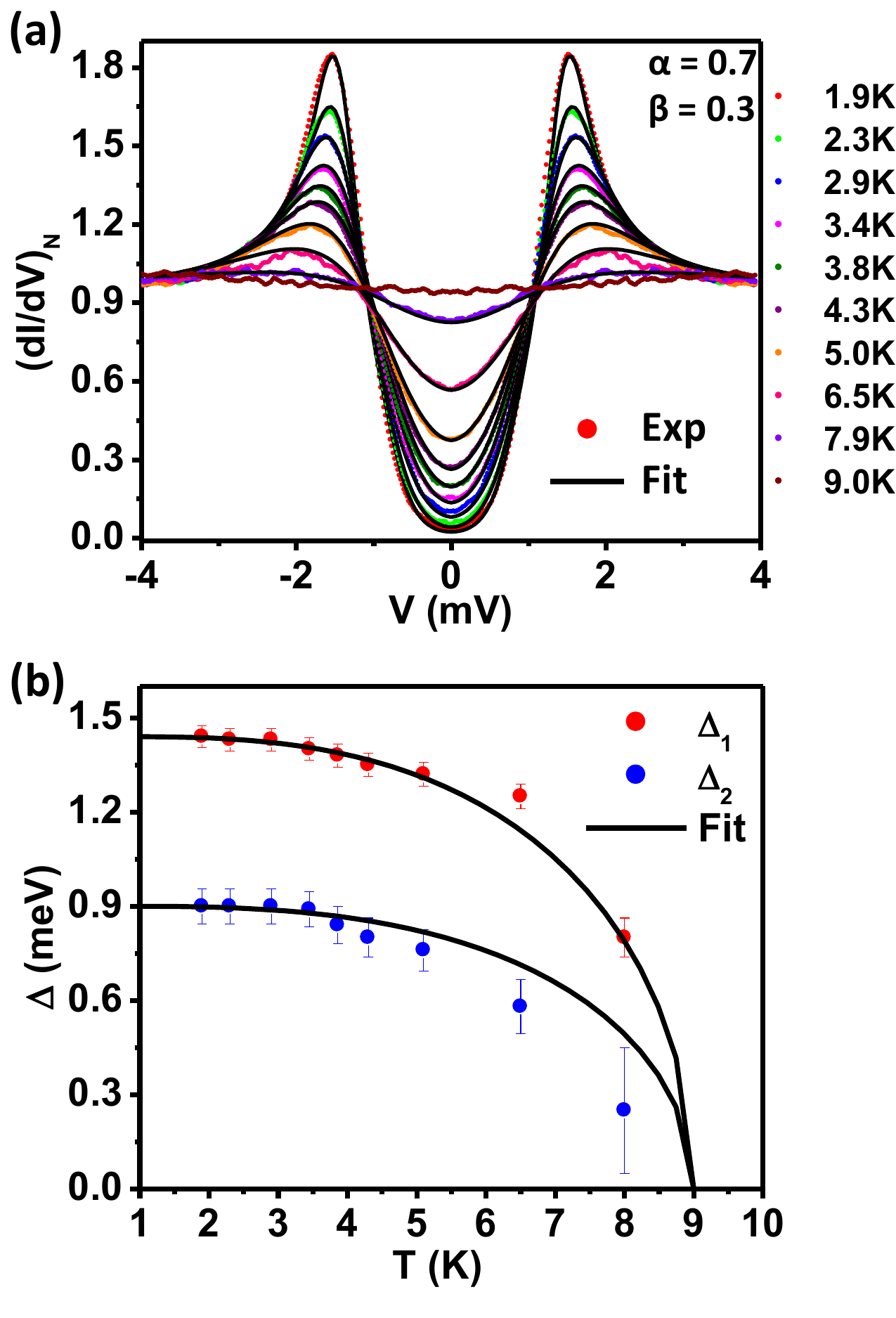}
	\caption{(a) Temperature dependence of tunneling conductance spectra (700mV, 250pA) with theoretical fits. The color dots are experimental data and black lines show fits with double gap. (b) Delta vs temperature plot extracted from plot (a), the dots are values extracted from the theoretical ﬁts and the solid line shows the temperature dependence as per BCS theory.}
	\label{Figure 4}
\end{figure}

\begin{figure}[h!]
	\centering
		\includegraphics[width=.44\textwidth]{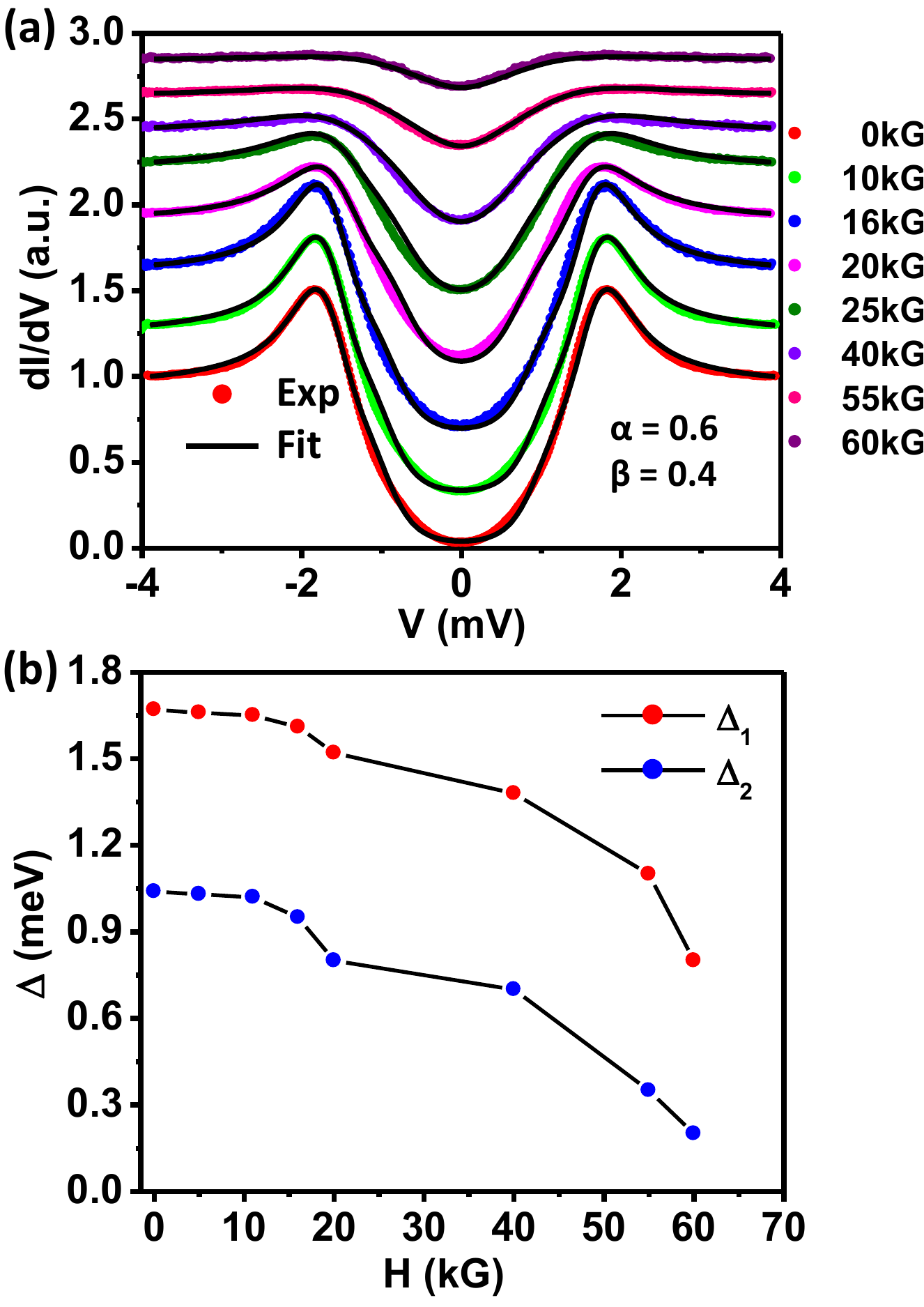}
	\caption{(a) Magnetic field dependence of tunneling conductance spectra (400mV, 250pA) with theoretical fits. The color dots are experimental data and black lines show fits with double gap. (b) Delta vs H plot extracted from plot (a).}
	\label{Figure 4}
\end{figure}
In Figure 1(a) we show an STM topographic image showing distinctly visible grains with average grain size $\sim$ 5 nm.  For spectroscopic measurements, we first brought the STM tip on the central parts of the grains and recorded the $dI/dV$ vs. $V$ spectra. We show three representative tunneling spectra in Figure 1(b,c,d). All the spectra are normalized to the conductance at 10 mV around which the conductance remains flat. In each of these panels we also show the best theoretical fits (green lines), first assuming a single gap originating from a single band in the Fermi surface.  For a single band superconductor, the tunneling current $I (V) \propto \int_{-\infty}^{+\infty} N_s(E)N_n(E-eV)[f(E) - f(E-eV)]dE$, where $N_s(E)$ and $N_n(E)$ are the normalized density of states of the BCS-like superconducting sample and the normal metallic tip respectively while $f(E)$ is the Fermi-Dirac distribution function.\cite{BCS} As per Dyne's formula $N_s(E) = Re\left(\frac{(E-i\Gamma)}{\sqrt{(E-i\Gamma)^2-\Delta^2}}\right)$, where $\Gamma$ is an effective broadening parameter included to take care of slight broadening of the BCS density of states possibly due to finite life time of quasiparticles.\cite{Dynes} This was used to calculate the single band $\frac{dI}{dV} = \frac{d}{dV}\left({G_N}\int_{-\infty}^{+\infty} N_s(E)N_n(E-eV)[f(E) - f(E-eV)]dE\right)$, where  $G_N = \frac{dI}{dV}|_{V>>\Delta/e}$. As it is clearly seen, the best theoretical fits within a single band model deviated significantly from the experimentally obtained spectra. We then attempted to fit the spectra within a simplistic two band model.\cite{2band model} If superconductivity appears in two distinct bands, then the tunneling current will have contributions from both the bands. Within a simplistic two-band model, the total tunneling current $I_{total} = \alpha I(V, \Delta_1, \Gamma_1) + \beta I(V, \Delta_2, \Gamma_2)$, where $\Delta_1$ and $\Delta_2$ are the gaps formed in the two different bands respectively and $\Gamma_1$ and $\Gamma_2$ are the corresponding effective broadening parameters. $\Gamma_1$ and $\Gamma_2$ also include the effective inter-band scattering, if any. The microscopic origin of $\Gamma$ in such analysis is of not much relevance and all physical processes leading to broadening are incorporated in $\Gamma$.  $\alpha$ and $\beta$ stand for the relative contribution of the two bands to the total tunneling current. Physically, $\alpha$ and $\beta$ could be associated with the crystal facet that the tip predominantly probes and how the crystallographic axis of a particular grain is oriented with respect to the tunneling barrier. $\alpha$ and $\beta$ might vary significantly when the tip moves from one particular orientation of a grain to another. As it is seen, the theoretically obtained spectra within the simplistic two band model fit remarkably well with the experimental spectra revealing the existence of two gaps with magnitude $\Delta_1 \sim$ 1.6 meV and $\Delta_2 \sim$ 0.9 meV respectively.

For all the three representative spectra presented in Figure 1 (b,c,d), $\alpha$ remained to be significantly larger than $\beta$. This observation is similar to the STM spectra obtained on polycrystalline MgB$_2$.\cite{Silva_MgB2} However, in case of MgB$_2$, for certain grains, spectral signature of the two gaps could be distinctly obtained on certain tunneling spectra recorded on appropriate grains where, due to particular orientation of the tunneling barrier with respect to the crystal plane of the grains, the smaller gap coupled more strongly than the larger gap. Motivated by this, we explored the tunneling spectra on a large number of grains and indeed for certain grains we achieved the ``two-gap" feature in a given tunneling spectrum. Two such representative spectra along with two-gap fits are shown in Figure 1(e) and Figure 1(f) respectively. While the gap amplitude for the smaller gap and the larger gap remained approximately same as before, for these spectra, $\alpha$ turned out to be smaller than $\beta$ which is consistent with the understanding that the two-gap feature is seen in a single spectrum when the band corresponding to the smaller gap has a larger contribution to the tunneling current. Therefore, based on the data presented above, we provided spectroscopic evidence of two-band superconductivity in Mo$_8$Ga$_{41}$. 

\begin{figure}[h!]
	\centering
		\includegraphics[width=.43\textwidth]{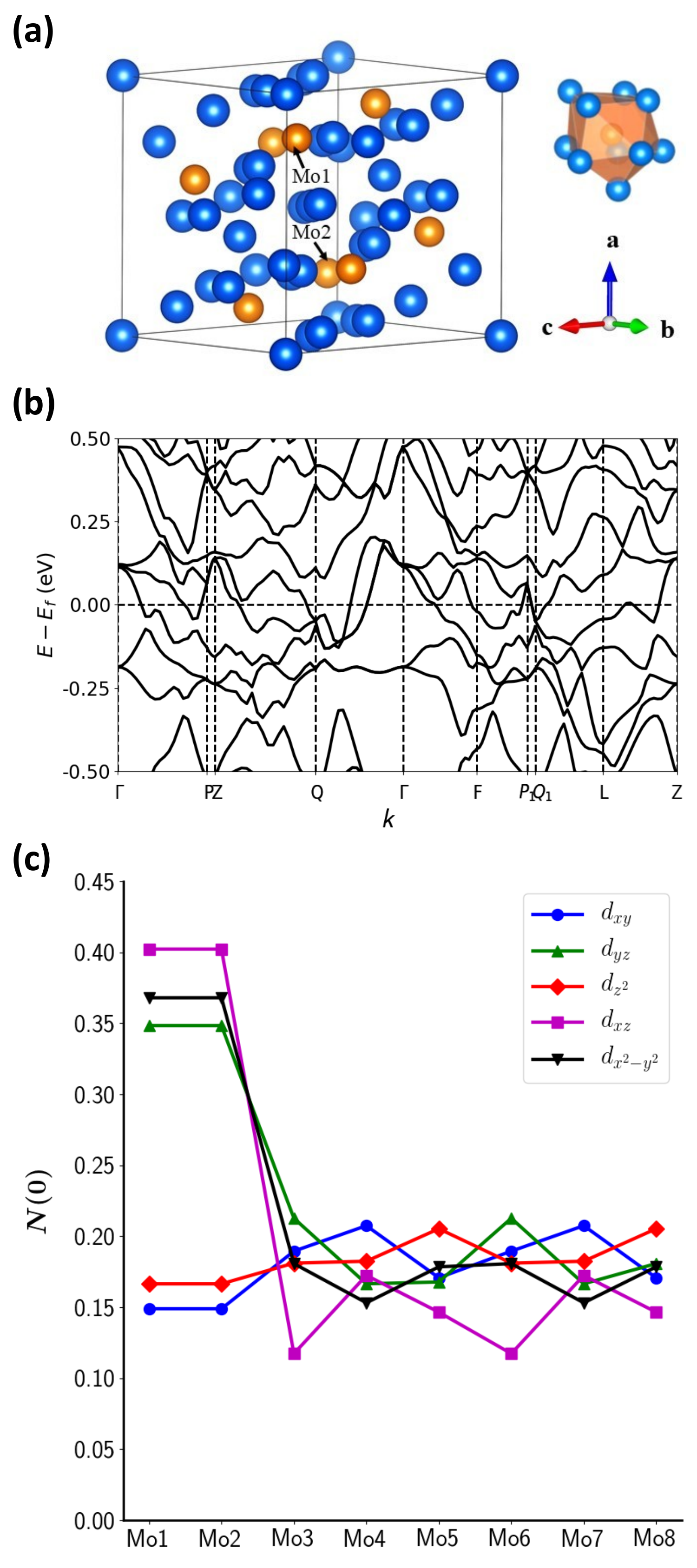}
	\caption{(a) Crystal structure of Mo$_8$Ga$_{41}$. (b) Computed band structure plotted along the high-symmetric directions. (c) Computed total density of states at the Fermi level for different Mo sites and different d-orbitals. The site-selectivity is clearly observed.}
	\label{Figure 4}
\end{figure}

To gain further understanding on the multiband superconductivity in Mo$_8$Ga$_{41}$ we now focus on the magnetic field dependence of the tunneling spectra. To understand the effect of the magnetic field on the superconducting energy gap, we fit the spectra recorded at different magnetic fields using the same formula that was used for the zero-field spectra. The spectra with fitting are shown in Figure 3(a). The extracted values of the two gaps as a function of magnetic field are shown in Figure 3(b). The larger gap ($\Delta_1$) decreases slowly and attains 53\% of its zero field value at a magnetic field of 6 Tesla, beyond which reasonable estimate of the gap was not possible. The smaller gap ($\Delta_2$), on the other hand, falls rapidly with increasing magnetic field. At a field of 6 Tesla, the gap becomes 0.15 meV which is only less than 15\% of the gap at zero field. This variation of the two gaps in Mo$_8$Ga$_{41}$ is similar to the variation of the two gaps with magnetic field in MgB$_2$, where the  small  gap is seen to disappear at  a magnetic field of approximately  1 Tesla  whereas  the  large  gap  remains almost unaffected within this range of magnetic field\cite{MgB2_5}. Furthermore, this observation is also consistent with the theoretical calculations of the vortex state of a multi-band superconductor with weak interband scattering\cite{Koshelev, Nakai}. For the larger gap of Mo$_8$Ga$_{41}$, 2$\Delta_1/k_BT_c$ is found to be 3.5 which is close to the expected value for a weak-coupling BCS superconductor. This suggests that the critical temperature $T_c$ in this compound is governed by the larger gap ($\Delta_1$). This again is similar to MgB$_2$ and YNi$_2$B$_2$C, where multiband scattering is weak\cite{YNBC_Goutam} and is again consistent with the theoretical expectation for a multiband superconductor with weak interband scattering\cite{Koshelev, Nakai}. Therefore, from our field dependent study of the superconducting energy gaps we surmise that the interband scattering in Mo$_8$Ga$_{41}$ is weak and falls in a range similar to that in MgB$_2$.  That might also explain why the qualitative spectral features in Mo$_8$Ga$_{41}$ are remarkably similar to those in MgB$_2$.


Now we focus on the nature of the two gaps of Mo$_8$Ga$_{41}$. In Figure 2(a), we show the temperature dependence of one representative spectrum over a temperature range from 1.9\,K to 9\,K. The colored dots represent the experimentally obtained spectra.  The coherence peaks gradually decrease with increasing temperature and the features associated with superconductivity disappear above 9\,K. In the same panel we also show the fits within the two-band model discussed above. For the entire temperature range, the values of $\alpha$ and $\beta$ remained fixed. The two gaps extracted (red and blue dots) are plotted with temperature in Figure 2(b). The black lines show the expected temperature dependence as per BCS theory\cite{BCS} for the two individual gaps with same $T_c$. As it is seen, the larger gap ($\Delta_1$) follows BCS temperature dependence. The smaller gap, on the other hand, remains constant up to almost 4 K and then starts gradually dropping and disappearing at 9 K showing only slight deviation from the BCS prediction.\cite{BCS} The disappearance of both the gaps at approximately the same temperature excludes the possibility of stoichiometric disorder in the grains on the sample. 

\begin{figure}[h!]
	\centering
		\includegraphics[width=.5\textwidth]{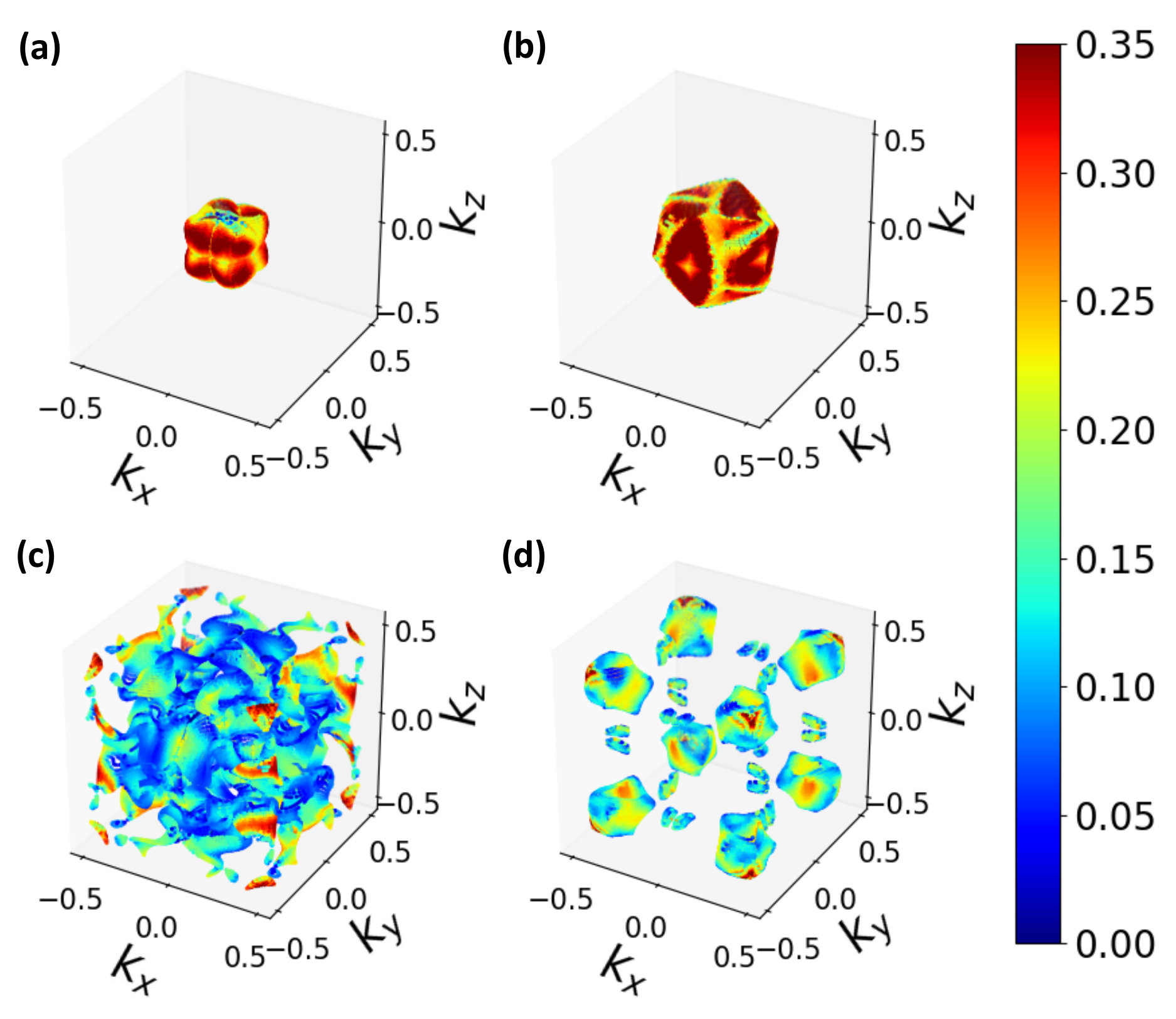}
	\caption{(a) 3D Fermi surfaces (pockets) with Fermi velocities shown as colour maps on the pockets.}
	\label{Figure 4}
\end{figure}

In order to understand the origin of exotic multiband superconductivity in Mo$_8$Ga$_{41}$, we investigated the band structure of the system through first-principles electronic structure calculations. The calculations are performed within the framework of density functional theory (DFT)\cite{DFT_1, DFT_2} using generalized gradient approximation (GGA)\cite{PBE} of the Perdew-Burke-Ernzerhof (PBE)\cite{PBE}form for the exchange-correlation functional as implemented in the Vinea Ab-initio Simulation Package (VASP). The projector augmented wave (PAW)~\cite{PAW} pseudo-potentials are used to describe the core electrons. Electronic wave-functions are expanded using plane waves up to a cut-off energy of 500 eV. The Monkhorst-Pack $k$-mesh is set to 6~$\times$~6~$\times$~6 in the Brillouin zone for the self-consistent calculation. All atoms are relaxed in each optimization cycle until atomic forces on each atom are smaller than 0.01 eV/\AA.

\color{black}

Mo$_8$Ga$_{41}$ belongs to the space group of $R\bar{3}$ (\# 148) with a rhombohedral structure.\cite{Muller} We obtained the relaxed lattice parameters as $a=b=c=9.5788\AA$, which are close to the experimental values, and $\alpha=\beta=\gamma=94.974^{o}$.\cite{Muller} Each Mo atom is surrounded by 10 Ga atoms, forming a polyhedron as shown in the Figure 4 (a).\cite{Cava} We discuss below that despite having the similar polyhedron nest around all 8 Mo atoms, two of them have stronger contributions to the Fermi surfaces. This is the primary origin of the site-selective superconductivity in this compound. To deal with the strong correlation effect of the $d$-electrons of the Mo atoms, we employed GGA+U method with $U=4$~eV. 

In Figure 4, we show one of our main theoretical findings. We evaluated site and orbital resolved density of states $N_{\sigma}(0)$ at the Fermi level, and compared them in Figure 4 (c). We notice that only Mo1 and Mo2 sites as shown in Figure 4 (a) contribute strongly to the Fermi surface. Among all the $d$-orbitals of the Mo atoms, $d_{xz}/d_{yz}$, and $d_{x^2-y^2}$ orbitals have strongest contributions, while the other orbitals and Mo atoms have significantly less contributions to the low-energy states. 

In Figure 4 (b), we show the band dispersion of the paramagnetic phase along the high-symmetric momenta directions. In order to identify the relation of the electronic structure with the observed multiple superconducting gaps, we now focus on the low-energy regime. We notice that there are four bands passing through the Fermi level with considerable three-dimensionality in all of them. 3D views of the corresponding Fermi surfaces are shown in Figure 5 with Fermi velocities plotted as a color map. We find two concentric hole pockets around the $\Gamma$-point, and one tiny electron pocket around the Brillouin zone corner. In addition, we also find a large and strongly anisotropic Fermi surface all over the Brillouin zone, a typical feature in this materials class. Within the BCS theory, superconducting order parameter is defined in the band basis in the \textbf{k}-space. Earlier,  in MgB$_2$, a two band superconductivity was reported where the inter-band electron-phonon coupling was found to play an important role.\cite{Silva_MgB2, MgB2_8} In superconducting iron-pnictide family, a multiband Fermi surface topology also leads to observation of multiple superconducting gaps.\cite{Rong_pnictide} By projecting the orbital weights of iron atoms onto the Fermi surface topology, recent reports found an exotic orbital selective characteristic in the superconducting order parameter. This orbital selective behavior provided important clues to the orbital fluctuations (or entangled  spin-orbital fluctuation owing to strong Hund's coupling in iron-pnictides) mediated superconducting pairing interaction. On the same footing, our observation of site-selective behavior on the low-energy electronic structure paves the way for a new mechanism of site-fluctuations induced pairing interaction responsible for superconductivity in Mo$_8$Ga$_{41}$.

Having shown the evidence of multi-band superconductivity in Mo$_8$Ga$_{41}$ we now attempt to identify the bands that might be responsible for the smaller and the larger superconducting gaps respectively. In order to identify that, we have calculated the effective Fermi velocities in different bands. The distribution of the Fermi velocity is shown as color maps on the four Fermi sheets as shown in Figure 5. It is found that the average velocity on the two pockets around the $\Gamma$-point is significantly larger than that around the other pockets. From qualitative understanding of phonon mediated pairing, it can be rationalized that for two distinct bands taking part in superconductivity, when the average Fermi velocity in a band is significantly larger than that in the other band, the superconducting gap forming on the band with higher average Fermi velocity should be lower than that forming in the other band. This is nothing but the manifestation of the semi-classical idea that it is harder for the faster electrons to interact strongly with the lattice as it spends less time near a given lattice point. On the other hand, slower electrons spend longer duration passing a given lattice points thereby leading to a larger electron-phonon coupling. This causes the band with relatively slower electrons form a stronger superconducting energy gap. A similar observation was made in case of the multiband superconductor YNi$_2$B$_2$C.\cite{YNBC_Goutam} Therefore, based on our data and theoretical analysis, we can conclude that the smaller gap forms in the bands shown in Figure 5(a or b) while the larger gap forms in the bands shown in Figure 5(c or d). Additional experiments like quantum oscillations in the superconducting state may provide information on the exact bands that participate in superconductivity. That might help identify the bands responsible for the two gaps with more precision.

In conclusion, we have provided direct spectroscopic evidence of multi-band superconductivity in the endohedral gallide Mo$_8$Ga$_{41}$ through detailed temperature and magnetic field dependent scanning tunneling spectroscopy and through band structure calculations. Analysis of the temperature dependent spectra within a two-gap BCS model revealed that both the gaps follow BCS temperature dependence. From a qualitative analysis of the magnetic field dependent STS spectra we surmise that in terms of the strength of interband scattering, Mo$_8$Ga$_{41}$ falls in the same range as MgB$_2$. Our band structure calculations revealed a unique site-selective mechanism that facilitates the observed multiband superconductivity in Mo$_8$Ga$_{41}$.  

GS would like to acknowledge financial support from the research grant of Swarnajayanti fellowship awarded by the Department of Science and Technology (DST), Govt. of India under the grant number DST/SJF/PSA-01/2015-16. SP acknowledges the SERB for support through grant number EMR/2016/003998.

\end{document}